# Modulation of Thermoelectric Power of Individual Carbon Nanotubes

Joshua P. Small, Kerstin M. Perez, and Philip Kim

Department of Physics, Columbia University, New York, NY 10027, USA.

**Abstract**

Thermoelectric power (TEP) of individual single walled carbon nanotubes (SWNTs) has been measured at mesoscopic scales using a microfabricated heater and thermometers. Gate electric field dependent TEP-modulation has been observed. The measured TEP of SWNTs is well correlated to the electrical conductance across the SWNT according to the Mott formula. At low temperatures, strong modulations of TEP were observed in the single electron conduction limit. In addition, semiconducting SWNTs exhibit large values of TEP due to the Schottky barriers at SWNT-metal junctions.

Enhanced thermoelectric phenomena in nanoscaled materials have attracted much attention recently owing to their potential for thermoelectric applications [1, 2]. In particular, the thermoelectric power (TEP) is of great interest in understanding transport, due to its extreme sensitivity to the change of electronic structure at the Fermi energy [3]. There have been a number of experimental efforts to measure the TEP of single walled carbon nanotubes (SWNTs) using macroscopic 'mat' samples [4-9]. These studies showed interesting thermoelectric phenomena, such as sign change of TEP upon absorption and desorption of gas molecules on the sample [6-8]. However, the numerous tube-tube junctions present in these macroscopic samples pose a problem in understanding the intrinsic thermoelectric properties of nanotubes. In addition, the ensemble average over different tube species makes it difficult to obtain the properties of an individual nanotube. Mesoscopic scale measurements able to probe individual nanotubes are necessary in order to elucidate their intrinsic thermoelectric properties.

In this letter, we present the results of mesoscopic TEP measurements on individual SWNTs. By employing microfabricated heaters and thermometers adjacent to a SWNT, we are able to measure the TEP at different values of the gate voltage. Remarkably, we find that the TEP of a SWNT can be strongly modulated by the gate voltage and is intimately related to the electrical conductance across the SWNT. Furthermore, we observe a greatly enhanced TEP in semiconducting SWNT field effect transistors (FETs), where the particle-hole symmetry is broken due to the presence of the Schottky barriers.

The inset to Fig. 1A displays a typical device used in our study for conductance and TEP measurements. The details of device fabrication will be published elsewhere [10]. In brief, SWNTs are grown on a silicon oxide/silicon (thickness of the oxide, $t_{ox}$ ~500 nm) substrate using chemical vapor deposition [11]. Thin Au/Cr metal contacts are fabricated on top of the SWNT by electron beam lithography. These lines serve as electrodes as well as thermometers. Another metal line placed near, but not in contact with the SWNT serves as a micro-heater. A current $I_h$ applied to the heater causes a temperature gradient to build up across the substrate via Joule heating. The temperature difference, $\Delta T$, between the two SWNT-electrode contacts is obtained by probing the changes in resistance, $\Delta R$, of the two metal stripes, $R_n$ and $R_f$. As $I_h$ changes, it is found that $\Delta R_n - \Delta R_f \propto \Delta T \propto I_h^2$ (Fig 1A), where the exact relations are obtained from the cross correlation between $R_n(T)$, $R_f(T)$ and $\Delta R(I_h)$ [12].

In our devices the underlying degenerately doped silicon substrate acts as a gate electrode to modulate the charge distribution of the SWNT. Fig. 1B shows the measured thermoelectric voltage, $\Delta V_p$, developed across the SWNT as a function of $I_h$ at various gate voltages, $V_g$ shown in Fig 1A. For fixed values of $V_g$ we observe that $\Delta V_p \propto I_h^2 \propto \Delta T$. The TEP of the SWNT device is obtained by the simple relation $S = \Delta V_p / \Delta T$. In addition to the thermoelectric measurement, the electrical conductance, $G$, of the same device is independently measured (Fig 1B inset). We find that both $G$ and the TEP are modulated by $V_g$ in most of the SWNT devices. Often, we observe sharp sign changes in $S$ (i.e., curvature changes in $\Delta V_p(I_h)$), while $G$ experiences only modest changes (i.e., a->b->c in Fig. 1B). It should be noted that the measured TEP contains contributions from both the absolute TEP of the SWNT, $S_{nt}$, and that of the metal electrode $S_m$, i.e., $S = S_{nt} - S_m$. Here, $S_m$ is expected to be insensitive to $V_g$, due to the bulk nature of the metal electrodes. Therefore, the variation of $S$ (as $V_g$ changes) can be solely attributed to that of $S_{nt}$. This observation clearly demonstrates that the TEP of a SWNT can be modulated simply by tuning the gate voltage.

The relationship between the conductance and the TEP can be further studied by comparing the two at the same $V_g$. Fig 2 shows a comparison between $G$ and $S$ of another metallic SWNT taken at temperatures $T = 300$ K and 100 K. The smooth variation of the conductance as a function of $V_g$ has been ascribed to resonant electron scattering by defects in metallic SWNTs, as studied in previous scanning gate microscopy experiments [13, 14]. These defects in SWNTs create multiple scattering sites in the conduction channel as $T$ decreases. As a result, $G$ develops a more complicated structure in $V_g$ at lower temperatures. Comparing $G$ and $S$, we discover an intimate relationship between the two quantities. It is found that after subtracting off a small offset from $S$, the sign of the observed TEP can be related with $dG/dV_g$ (see dashed lines in the figure). This offset is ~ 4 $\mu$V/K at $T = 300$ K, and it becomes smaller at lower temperatures. We believe that this small offset corresponds to $S_m$, which presumably decreases in magnitude along with $T$ [15]. Thus, across the extrema of $G$, $S_{nt}$ changes its sign as $V_g$ changes.

A quantitative correlation between $S$ and $G$ can be made by the comparing the measured TEP to that predicted by the Mott formula, which relates the diffusive thermopower, $S_d$, to the conductance [16]:

$$S_d = \frac{-\pi^2 k_B^2 T}{3|e|} \frac{1}{G} \frac{dG}{dE}\bigg|_{E_f} \qquad (1)$$

where $E_F$ is the Fermi energy. Although this formula was originally derived for bulk systems, it is still valid in 1-dimensional (1-d) mesoscopic conducting channels, where we can incorporate the Landauer expression, $G = (2e^2/h)t$, with the electron transmission coefficient, $t$ [17]. Thus, $S_d$ is estimated from the experimentally obtained $G$ vs. $V_g$ by relating the change in electrochemical potential of the channel, $\Delta E_F$, with the change in applied gate voltage, $\Delta V_g$. At high temperatures, where the Coulomb charging effect is negligible, $\Delta E_F = h v_F C_g \Delta V_g / 8e$, where the Fermi velocity $v_F = 8.5 \times 10^5$ m/sec, and $C_g$ is the capacitive gate coupling per unit length [18]. $C_g = 15$ pF/m is estimated from Coulomb blockade analysis, in reasonable agreement with geometric considerations ($t_{ox}$ ~500 nm). The calculated thermopower from eq. (1) (dotted lines in low panels of Fig. 2) agrees reasonably well with the measured TEP. We found similar agreement between $S$ and $S_d$ calculated from the measured conductance in more than 5 metallic SWNT devices measured in the temperature range $T > \sim 50$ K [19].

We now turn to the effect of temperature on TEP and $G$. In particular, we will focus on the low temperature regime, where single electron tunneling is the dominant transport mechanism. Fig. 3 displays $G$ and TEP vs. $V_g$ at various temperatures. The observed periodic oscillations of $G$, apparent at $T < 30$ K, are a clear signature of single-electron charging effects. The charging energy $E_C = e^2 / 2C = 6$ meV, where $C$ is the self capacitance of the SWNT, is obtained from standard analysis of linear and nonlinear transport in the Coulomb blockade (CB) regime [20] (inset of Fig. 3A). We find that the observed TEP exhibits periodic oscillations as a function of $V_g$ in the CB regime (Fig. 3B). The TEP in the CB regime has been studied in semiconductor quantum dots in a 2-dimensional electron gas [21-25]. Gate induced TEP oscillations similar to those we observe in SWNTs were reported in these previous studies. However, in these quantum dot experiments the absolute value of the TEP was obtained from an indirect estimation of the temperature gradient across the sample. In our experiments, a direct measurement of $\Delta T$ allows us to quantitatively compare experimental observations with theoretical models. We find the amplitude of the TEP oscillation in the SWNT increases as temperature decreases, and the averaged amplitude scales as $S_{rms} \approx (k_B/2e) E_C/k_B T$ (inset of Fig. 3B). Similar scaling behavior of TEP was observed in other metallic SWNTs at low temperatures [26].

Finally, we discuss the TEP of semiconducting SWNTs. Fig. 4 displays $G$ and TEP for a typical semiconducting SWNT [27]. Three distinct regions of interest are defined by the threshold voltage $V_{th}$: (i) $V_g \ll V_{th}$, $G$ is large and decreases slowly as $V_g$ increases; (ii) $V_g \sim V_{th}$, $G$ decreases exponentially as $V_g$ increases; (iii) $V_g \gg V_{th}$, $G$ is strongly suppressed. For a semiconducting SWNT weakly coupled to the electrodes, this field effect behavior has been ascribed to Schottky barrier (SB) controlled transport [28, 29]. The inset of Fig 4 shows schematic potential diagrams with the two SBs formed at each end of the SWNT. Electronic conduction is dominated by tunneling through the SB in region (i), and by thermionic conduction over the SB in region (ii). We find that the measured TEP exhibits distinctly different behavior in each region (lower panel). The TEP saturates at a constant value of $S \sim 120 \, \mu$V/K in (i), and exhibits a peak at a value of $S \sim 260 \, \mu$V/K in (ii) at room temperature.

This large modulation of the TEP in semiconducting SWNTs can be understood by considering the broken particle-hole symmetry at the SB as follows. For the region $V_g \ll V_{th}$, the WKB method gives $G(E) \approx \frac{4e}{h} e^{-\alpha\sqrt{\Delta+E}}$, assuming that the transport is dominated by charge tunneling through the SB barrier of height $\Delta$, where $E$ is measured from $E_F$. Here $\alpha$ is a constant independent of $E$, since the SB width is saturated to $\sim t_{ox}$ when $V_g \ll V_{th}$ [30]. Thus, from eq. (1), we expect the TEP to saturate at a value $S = \frac{\pi^2 k_B^2 T}{6|e|\sqrt{\Delta}} \ln \frac{G(0)}{4e^2/h}$, where $G(0)$ is obtained from the value of the saturated conductance measured in region (i). From the measured

TEP and $G$ in this regime, we obtain $\Delta = 160$ meV, which agrees well with the observed *p*-type behavior in this SB SWNT FET. On the other hand, near the subthreshold regime, $V_g \sim V_{th}$, thermionic conduction over the SB is the most important transport mechanism, and $G(E) \propto e^{-(E-\Delta)/k_B T}$ [31]. As a result, we expect to find positive values of the TEP as large as $S = \pi^2 k_B / 3|e| = 284 \, \mu$V/K, which agrees well with the observed peak value.

In conclusion, we report gate field modulation of TEP of individual SWNTs. These results show that mesoscopic TEP provides an independent experimental tool extremely sensitive to electron-hole asymmetry in 1-d conduction channels in SWNTs. In particular, the ability to modulate the TEP as demonstrated in this work opens up the possibility of using SWNTs for thermoelectric applications. Greatly enhanced values of the TEP discovered in semiconducting SWNT devices show potential promise for thermoelectric cooling and generation using these materials.

We thank M. Sfeir, K. Bosnick, and L. Brus for help with nanotube growth, and J. Hone and I. Aleiner for helpful discussion. This work is supported primarily by the Nanoscale Science and Engineering Initiative of the National Science Foundation under NSF Award Number CHE-011752.

[26] It is noted that the amplitude of TEP oscillation is smaller than that expected from a simple non-interacting electron model [24] by a factor of ~2, and that the peak shape of TEP exhibits a more complicated structure at $k_B T \ll E_C$. A detailed study is in progress.

[27] For a typical semiconducting SWNT device with Au/Cr contacts, most devices as-made exhibit *p*-type behavior. Although *n*-type or ambipolar SWNT FET devices were often obtained after annealing the devices, we focus only *p*-type devices in this work.

**Figure Captions**

**Fig. 1**. (A) The inset shows a scanning electron micrograph of a typical SWNT device used in this study. Two stripe electrodes, $R_n$ and $R_f$, contacting the SWNT, are used as thermometers to probe junction temperatures by measuring their four-probe resistance. The graph shows the normalized change in $R_n$ and $R_f$ as a function of the heater current, $I_h$ at $T = 300\,\text{K}$. Changes in resistance are mapped to the change in temperature difference, $\Delta T$, across the SWNT. (B) The thermoelectric voltage measured as a function of $I_h$ at $T = 300\,\text{K}$. Dashed (a), dotted (b), and solid (c) lines correspond to data taken at various applied gate voltages marked by arrows in the inset. The inset shows the conductance of the SWNT as a function of gate voltage. The conductance is measured by applying 100 $\mu$V of excitation voltage using a lock-in amplifier.

**Fig. 2**. Measured conductance and TEP (solid line) vs. applied gate voltage at $T = 300\,\text{K}$ (A) and 100 K (B). Vertical dashed lines correspond to values of gate voltage where the conductance is extremal. Horizontal dashed lines indicate the offset of TEP matching the small TEP of the metal electrodes. The dotted lines in the lower panel represent the Mott formula calculated from the measured conductance (see text).

**Fig. 3**. (A) Conductance of a metallic SWNT as a function of gate voltage taken at various temperatures. The inset shows the conductance at finite source drain bias voltage ($V_{sd}$) in gray scale. The white and black color represents 0 and 3 $\mu$S, respectively. (B) The corresponding TEP of the SWNT device in (A). The graphs at different temperatures are shifted for clarity. Dashed lines represent $S = 0$ for each data set. The inset shows the root-mean-squared TEP ($S_{rms}$) calculated from the data set at different temperatures. The dotted line corresponds to a simple fit to $1/T$.

**Fig.4**. Conductance (upper panel) and TEP (lower panel) versus gate voltage of a semiconducting SWNT device at $T = 300$ K. The conductance plot exhibits typical *p*-type SWNT FET behavior, where the conductance is suppressed by more than two orders of magnitude around the threshold voltage $V_{th}$. Large values of TEP are observed near the threshold voltage. The inset shows the band alignment of the SB model in different regions of the gate voltage.

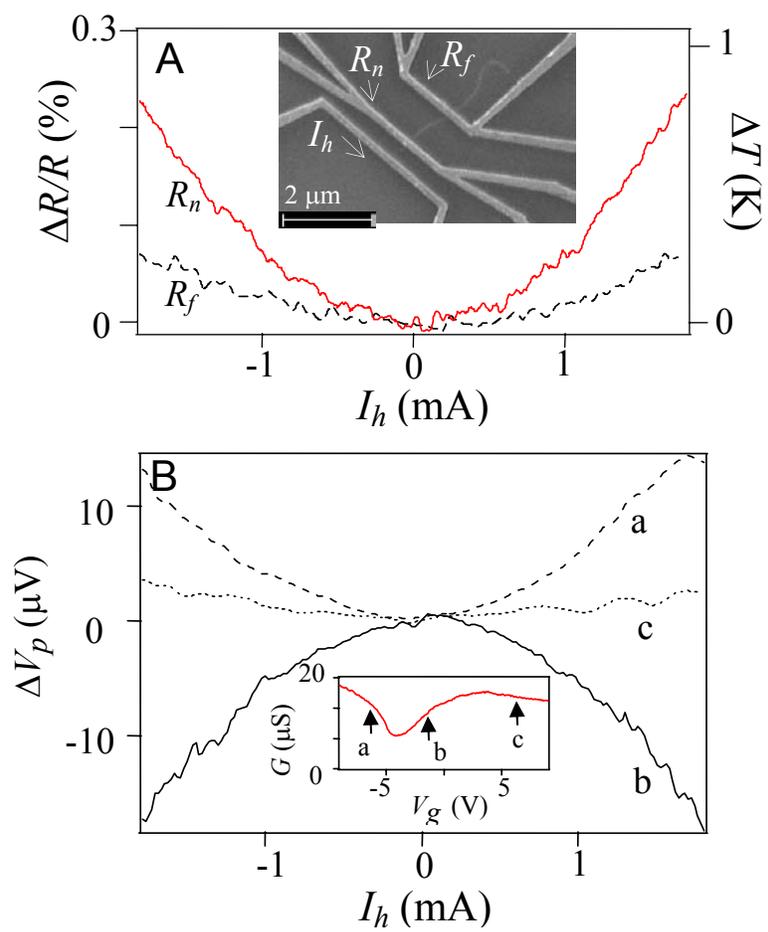

Fig 1.

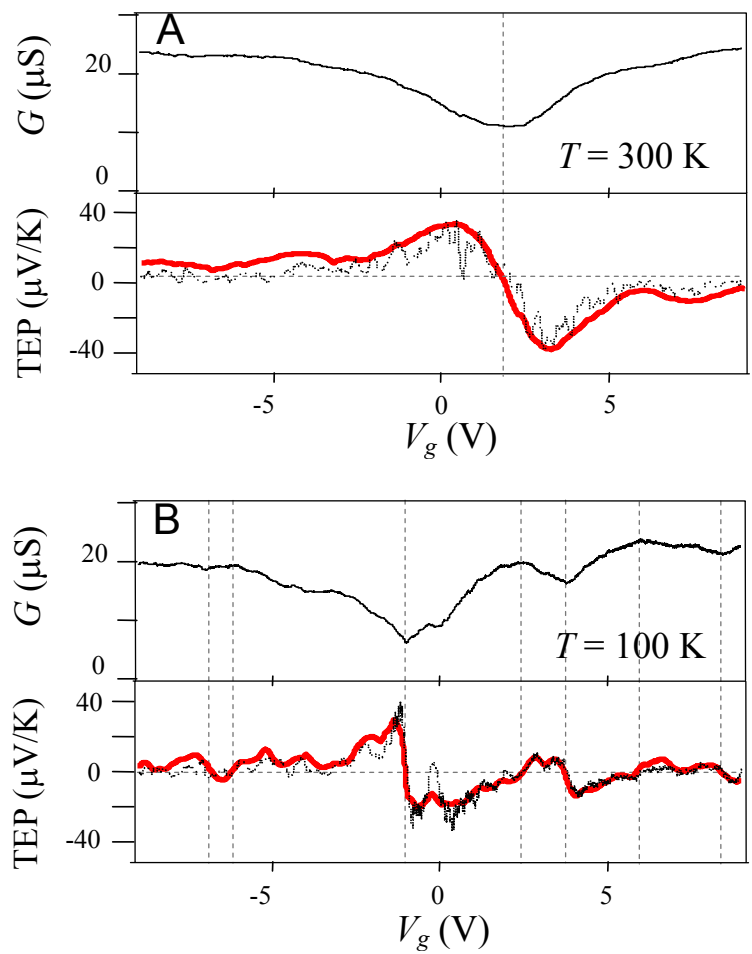

Fig 2.

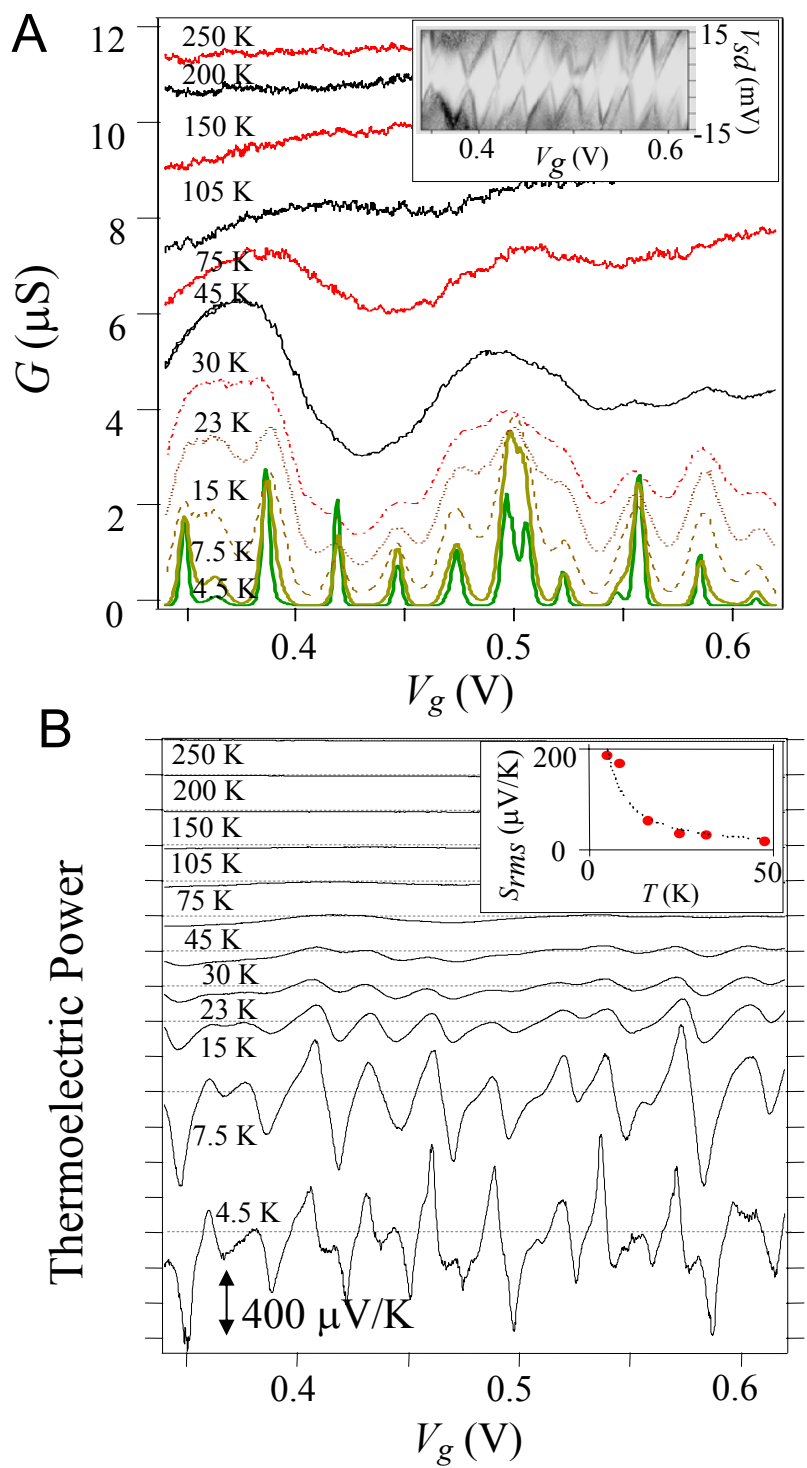

Fig 3.

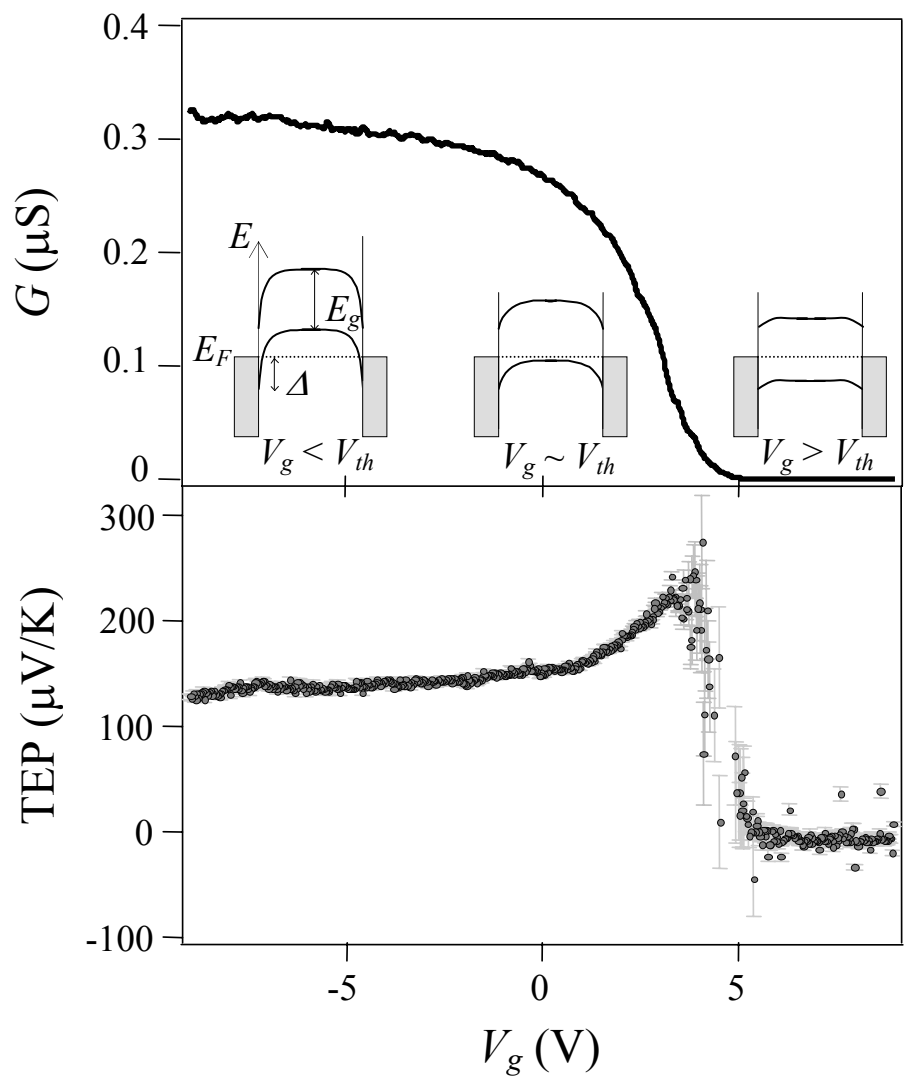

Fig 4.